\newcommand{\kl}[3]{\mbox{$\rm #1$}^{\mu\nu , \alpha\beta}_{#2}(#3)}

\def\be{\begin{equation}}
\def\te{\end{equation}}
\def\bea{\begin{eqnarray}}

\def\tea{\end{eqnarray}}

\def\k{\kappa}

\def\m{\mu}
\def\n{\nu}

\newskip\humongous \humongous=0pt plus 1000pt minus 1000pt

\newif\ifdtup

\def\ha{{1\over 2}}

\documentstyle[eqsecnum,prd,aps]{revtex}

\makeatletter                    
\@addtoreset{equation}{section}  
\makeatother                     

\begin{document}
\title{Quantum Noise and Fluctuations in Gravitation and Cosmology}
\author{B. L. Hu$^a$\thanks{Corresponding author. Electronic address:
hub@physics.umd.edu}, Albert Roura$^a$,
Sukanya Sinha$^b$, E. Verdaguer$^c$
\\
$^a${\small Department of Physics, University of Maryland,
             College Park, Maryland 20742-4111, U.S.A.}\\
$^b${\small Indian Statistical Institute, Bangalore Centre,
 8th Mile, Mysore Road, Bangalore-560059, INDIA}\\
$^c${\small  Departament de Fisica Fonamental and C.E.R. in
Astrophysics, Particles and Cosmology, Universitat de Barcelona,
Av.~Diagonal 647, 08028 Barcelona, Spain}}

\date{(\small UMD Physics Preprint number: umdpp 03-045, April 12, 2003)}
\maketitle \centerline{\it Invited Talk given by BLH at the First
International Symposium on Fluctuations and Noise}
\centerline{\it Sponsored by SPIE, 1-4 June 2003, Santa Fe, New
Mexico. Paper number 5111-46}
\begin{abstract}
We give a short update of our research program on nonequilibrium
statistical field theory applied to quantum processes in the early
universe and black holes, as well as the development of
stochastic gravity theory as an extension of semiclassical
gravity and an intermediary in the 'bottom-up' approach to quantum
gravity.
\end{abstract}

\section*{Prologue} (BLH) "It is no secret that the message
of this conference is `NOISE IS GOOD'. In this talk I want to show
that {\it not only is noise good, it is absolutely essential}...
" . These words were not uttered in this conference, as it would
have been acausal, but ten years ago, at a  workshop devoted to
the nascent yet fascinating subject of Fluctuations and Order
\cite{HMLA}. After a decade, the tenor and objectives of this
statement are even closer to reality, as witnessed by the strong
focus and great variety of this symposium.

I substantiated this claim then by an enumeration of the many
processes in gravitation and cosmology where quantum noise and
fluctuations play an active role:

\noindent 1. Particle creation as parametric amplification of
vacuum fluctuations\\
2. Thermal radiance from accelerated observers and black holes
   as fluctuation-dissipation phenomena\\
3. Entropy generation from quantum stochastic and kinetic processes\\
4. Phase transitions in the early universe as noise-induced processes\\
5. Galaxy formation from primordial quantum fluctuations\\
6. Anisotropy dissipation from particle creation as backreaction processes\\
7. Dissipation in quantum cosmology and the issue of the initial state\\
8. Decoherence, backreaction  and the semiclassical limit of quantum gravity\\
9. Stochastic spacetime and continuum limit, gravity as an effective theory\\
10. Topology change in spacetime and loss of quantum coherence problems\\
11. Gravitational entropy, singularity  and time asymmetry\\
12. `Birth' of the universe as a spacetime fluctuation and
tunneling phenomenon\\

The above list was prepared for cosmological issues. Further
developments in the last ten years focusing on the effects of
noise and fluctuations using the statistical field theory
approach we have developed (for a review, see \cite{Banff,CHM})
include, related to Topic 1 : Preheating in post-inflationary
cosmology \cite{RH1,RH2}; Topic 2: Thermal and near-thermal
radiance in detectors, moving mirrors, black holes and cosmology
\cite{RHA,RHK,KHMR,KMH}; Topic 3: Correlation entropy
\cite{CH00,CHcorent}; Topic 4: Defect formation \cite{SCHR},
tunneling induced by quantum and thermal noise fluctuations
\cite{CRVopensys,CRVtunnel}. Topic 5 is presently under pursuit
\cite{RouVer03a} and partially summarized in the last section of
this paper. Topic 9 includes preliminary work on wave propagation
in stochastic spacetimes \cite{HuShi} and mesoscopic fluctuations
\cite{Tomeso}.

In  black hole physics, focusing again on fluctuations, one can
mention topics on  black hole fluctuations and backreaction
\cite{HRS,SRH} (see references therein for other related work) and
the energetics and dynamics of black hole phase transition
\cite{GregBH,GPY,WhiYor}. The wish list would include:
applications of stochastic gravity to the statistical mechanical
definition of entropy and statistical field theory of black hole
nonequilibrium thermodynamics.

Using statistical quantum field theory , Topics 1, 2, 3, 6 are
quite well understood, 7, 8 only partially, and Topics 4, 5, 9
are currently under pursuit. Certain concepts in Topics 10-12 may
not even be well-defined (e.g., what does the `birth' of the
universe mean?), but the newly established theory of stochastic
gravity (for a review, see \cite{stogra,HVErice,CQGrev,LivRev},
an ongoing development, see \cite{RouVer03b}) may offer
alternative ways to address these issues, as well as provide an
intermediary towards quantum gravity (our definition is different
from the ordinary, see \cite{KinQG}).

In this talk I will focus on the theory of noise in quantum
fields and how quantum fluctuations could have played an active
and even decisive role in many fundamental processes in cosmology
and gravitation, especially near the Planck time ($10^{-43}$ sec
from the Big Bang). I will describe how stochastic gravity theory
can be understood easily from an open system conceptual framework,
and give two examples of its applications: fluctuations and
backreaction in black holes, and structure formation in the early
universe.

The Planck time is the time when many familiar features of
spacetime depicted by Einstein's theory of general relativity
give way to an as-yet-unknown quantum theory of gravity depicting
the microstructure of spacetime. (Some theoreticians believe the
superstring theory is the answer.) Just below the Planck energy
we believe the universe can be adequately described by a
semiclassical theory of gravity \cite{DeW75,BirDav,scg}, where
quantized matter fields coexist with a classical spacetime. Many
qualitative changes are believed to have taken place at this
energy scale, amongst them the formation of spacetime depictable
as a manifold, the emergence of time, the creation of particle
pairs from the vacuum, the growth of fluctuations as seeds for
galaxies, and possible phase transitions and the ensuing entropy
generation processes. It is also the cross-over point of quantum
to classical and micro to macroscopic transitions.\footnote{It is
for this reason that I think one can  view {\it general relativity
as geometro-hydrodynamics} \cite{GRhydro}, {\it semiclassical
gravity as a mesoscopic physics} \cite{meso}, and take a {\it
kinetic theory approach to quantum gravity}\cite{KinQG}. The only
difference is that instead of dealing with the quantum to
classical and micro- to macro- transition in the state of matter
and fields we are dealing with the corresponding issues for
spacetime and geometry.}

In Section I, we first explain the origin and nature of noise in
quantum systems interacting with an environment, using the
influence functional method. (For non-Ohmic bath at low
temperatures, colored noise and nonlocal dissipation would
appear;  and for nonlinear coupling multiplicative noise is
generally expected. A generalized fluctuation-dissipation
relation for these systems can be proven, and the stochastic
(master, Langevin and Fokker-Planck) equations derived, depicting
the dissipative dynamics of the open system under the influence
of noise. We then discuss stochastic gravity theory in Sec. 2 and
the two examples in Sections 3 and 4.

\section{Quantum Fluctuations in Open Systems}


We begin by describing how the concepts of quantum open systems
can be of use in the treatment of statistical mechanical problems
involving quantum fields. We start with two subsystems A and B.
When the precise information of subsystem B is not required, but
only its averaged effect on subsystem A is of interest, one can
{\it coarse-grain} B and include its averaged effect on A, which
involves finding the {\it backreaction} on A.
In so doing  A is rendered an open system, with B acting as its
environment.
For the analysis of open systems with backreaction from the
environments the influence functional (IF) formalism of Feynman
and Vernon {\cite{if,qbm}} proves useful.
Let us first use a simple example from quantum mechanics to
illustrate the idea and the method. We then show how we can use
this framework to address issues in gravitation and cosmology.

\subsection{Stochastic Effective Action in Quantum Open Systems}
\label{sec:open} We begin with a brief schematic summary of the IF
formalism as applied to a simple system. Consider a system $S$,
described by the degrees of freedom $x$, interacting with an
environment $E$, described by the degrees of freedom
$q$.\footnote{We are labeling the degrees of freedom of the
system and the environment by single letters $x$ and $q$ with the
understanding that they can represent multiple or even infinite
degrees of freedom, e.g. corresponding to a field \cite{HM2}.} The
full closed quantum system $ S + E$ is described by a density
matrix $\rho (x,q;x',q',t)$. If we are interested only in the
state of the system as influenced by the overall effect, but not
the precise state of the environment, i.e, the dynamics of the
open system, then the reduced density matrix
$\rho_r(x,x',t)=\int~dq~\rho (x,q;x',q,t)$ would provide the
relevant information. (The subscript $r$ stands for reduced.)
Assuming that the action of the coupled system decomposes as
$S=S_s[x]+S_e[q]+S_{int}[x,q]$, and that the initial density
matrix factorizes (i.e., takes the tensor product form), $\rho
(x,q;x',q',t_i)=\rho_s(x,x',t_i)\rho_e(q,q',t_i)$, the reduced
density matrix is given by \be \rho_r(x,x',t)
=\int\limits_{-\infty}^{+\infty}dx_i\int\limits_{-\infty}^{+\infty}dx'_i~
\int_{x_i}^{x_f} Dx~ \int_{x'_i}^{x'_f} Dx '~
e^{i(S_s[x]-S_s[x']+S_{IF}[x,x',t])} ~\rho_r(x_i,x'_i,t_i~)
\label{pathint} \te where $S_{IF}$  is the influence action
related to the influence functional $\cal F$ defined by \be {\cal
F}[x, x'] \equiv e^{iS_{IF}[x,x',t]}\equiv\int~dq_f~dq_i~ dq_i'~
  \int_{q_i}^{q_f}Dq~\int_{q'_i}^{q_f}Dq'~
e^{i(S_e[q]+S_{int}[x ,q]-S_e[q']-S_{int}[x ',q'])}
\rho_e(q_i,q'_i,t_i). \label{SIF} \te $S_{IF}$ in general is complex.
Retaining only quadratic terms (an approximation which covers many of
the interesting applications that we will consider later), we may
write \be S_{IF}(x,x')=\int~dt~dt'~\{ {1\over
2}(x-x')(t)D(t,t')(x+x')(t') +{i\over 2}(x-x')(t)N(t,t')(x-x')(t')\}
\label{SIFquad} \te where $D$ and $N$ stand for the real dissipation
and noise kernels respectively. Note that in this quadratic order
approximation, the influence action $ S_{IF}(x,x')$ is related to the
closed-time-path(CTP) or in-in effective action (for details on the
CTP effective action see \cite{ctp})  $\Gamma_{CTP}[x,x']$ through
\be \Gamma_{CTP}[x,x'] = S[x] - S[x'] + S_{IF}[x,x'].
\label{CTPIFconn} \te The equation of motion obtained from the CTP
effective action for the expectation values is clearly seen to be
real and causal \cite{ctp}. It reads \be \left.{\delta\over \delta
x(t)}\Gamma_{CTP}[x, x']\right|_{x' = x = {\bar x}} = 0 \label{CTPem}
\te From the influence functional a Langevin equation for the system
dynamics may be derived by a formal procedure, first introduced by
Feynman and Vernon \cite{if}, which consists of introducing  a
Gaussian stochastic source $\xi(t)$ with $\langle\xi(t)\rangle_{\xi}
= 0$ and $\langle\xi(t)\xi(t')\langle = N(t,t')$ and defining an
improved or stochastic effective action as \be S_{eff}[x,x';\xi] =
S_s[x] - S_s[x'] + {\cal{R}}S_{IF}[x,x'] + \xi (x-x') \label{Seffxi}
\te such that $\left < e^{iS_{eff}[x,x';\xi]}\right >_{\xi} =
e^{i\Gamma_{CTP}[x,x']}$. This leads to equations of motion with a
stochastic force: \be \left.{\delta S_{eff}[x,x';\xi]\over \delta
x}\right|_{x =x'} = 0 \quad \mathrm{or, equivalently,} \quad
\left.{\delta \Gamma_{CTP}[x,x';\xi]\over \delta x}\right|_{x =x'} =
0.
 \label{stochem} \te
The equation of motion obtained from (\ref{stochem}) using
(\ref{Seffxi}) is  \be {\partial S_s\over\partial
x(t)}+\int~dt'~\gamma (t,t') {dx(t')\over dt'}= \xi
\label{langevin} \te where $D(t,t')=-\partial_{t'}\gamma (t,t')$.
Being now in the form of a Langevin equation, the physical meaning of
the $\gamma$ and $N$ kernels in Eq. (\ref{langevin}) becomes
clearer. Both the terms involving $\gamma$ and $\xi$ represent the
backreaction of the environment on the system.  However, $\gamma$ (or
more properly the odd part of $\gamma$) is associated with dissipation
and $\xi$ is a stochastic noise term associated with random
fluctuations of the system, exactly as the terms are interpreted in
the context of Brownian motion. Averaging (\ref{langevin}) over the
noise using the appropriate probability distribution will give the
semiclassical equation of motion for the mean value of $x$. Since the
noise and dissipation arise by considering a subsystem within a closed
system (as is done here, as opposed to being put in by hand), they are
in general related by a set of generalized fluctuation-dissipation
relations (FDR), which can be represented by a linear, non-local
relation of the following form, provided that the Hamiltonian for the
environment and the system-environment interaction are time
independent and the initial state of the environment is stationary
with respect to the Hamiltonian of the environment: \be N(t-t') =
~\int~d(s -s')K(t-t',s-s')\gamma(s -s') \label{FDR} \te To keep the
discussion simple, we have written the noise and dissipation kernels
in terms of single scalar functions. However, the method is general
enough to encompass multiple noise and dissipation kernels and cases
where the kernels are tensorial, as in the stochastic gravity theory
discussed later.



\section{Stochastic Gravity and Metric Fluctuations}
\label{sec:stogra}


Stochastic semiclassical gravity \cite{ELE} of the 90's is a
theory naturally evolved  from semiclassical gravity \cite{scg} of
the 80's and quantum field theory in curved spacetimes
\cite{DeW75,BirDav} of the 70's. Whereas semiclassical gravity is
based on the semiclassical Einstein equation with sources given
by the expectation value of the stress-energy tensor of quantum
fields, stochastic semiclassical gravity is based on the
Einstein-Langevin equation, which has in addition stochastic sources
with correlation functions characterized by the noise kernel.
the noise kernel. The noise kernel is the vacuum expectation
value of the  (operator-valued) stress-energy bi-tensor which
describes the fluctuations of quantum matter fields in curved
spacetimes.

\subsection{From Semiclassical to Stochastic Gravity } \label{sec1}

The first stage in the road to stochastic gravity begins with {\it
quantum field theory in curved spacetime}, which describes the
behavior of quantum matter fields propagating in a specified (not
dynamically determined by the quantum matter field as a source)
background gravitational field. For a scalar field $\phi$ it obeys
the wave equation $ (\Box + m^2) \phi(x) = 0 $ where $\Box$ is the
Laplace-Beltrami operator,  which contains the imprint of the
curvature of the background spacetime. In this theory the
gravitational field is given by the classical spacetime metric
determined from classical sources by the classical Einstein
equations, and the quantum fields propagate as test fields in such a
spacetime. For time dependent spacetime geometry it may not be
possible to define a physically meaningful vacuum state for the
quantum field at all times. Assuming that one defines a vacuum state
at some initial time, the vacuum state at a latter time will differ
from that defined initially because particles are created in the
intervening time. An important process described by quantum field
theory in curved spacetime is indeed particle creation from the
vacuum (and effects of vacuum fluctuations and polarizations) in the
early universe \cite{Hawking} and Hawking radiation in black holes
\cite{Hawking}.

The second stage in the description of the interaction of gravity
with quantum fields is {\it back-reaction}, i.e., the effect of
quantum fields on the spacetime geometry.
The dynamic classical spacetime metric creates particles of the
quantum field and these in turn provide a backreaction on the
spacetime metric which alters its dynamics in response. One
assumes a general class of spacetime where the quantum fields
live in and act on, and seek a solution which satisfies
simultaneously the Einstein equation for the spacetime and the
field equations for the quantum fields. The Einstein equation
which has the expectation value of the stress-energy operator of
the quantum matter field as the source is known as the {\it
semiclassical Einstein equation}:
 \be \label{semi} G_{\m\n} (g_{\alpha \beta})
 = \k \langle \hat{T}_{\m\n} \rangle_q
 \te where $\hat{T}_{\m\n} $ is the
stress-energy tensor operator of, say, a free scalar field $\phi$,
$G_{\mu\nu}$ is the Einstein tensor, $\kappa = 8\pi G_N$ and $G_N$ is
Newton's constant. Here $\langle\,\rangle_q$ denotes the expectation
value taken with respect to some quantum state compatible with the
symmetries of the background spacetime, a classical object.
The theory obtained from a self-consistent solution of the
geometry of the spacetime and the quantum field is known as {\it
semiclassical gravity}. Incorporating the backreaction of the
quantum matter field on the spacetime is thus the central task in
semiclassical gravity.

Studies of the semiclassical Einstein equation for the
backreaction problems have been carried out in the last two
decades by many authors for cosmological and black hole
spacetimes. A well-known example of semiclassical gravity is the
damping of anisotropy in Bianchi universes (which is the basis of
chaotic cosmology in the 70's) by the backreaction of vacuum
particle creation, and inflationary cosmology
\cite{infcos,KolTur90,Linde90} of the 80's driven by a constant
vacuum energy density source such as the expectation value of a
Higgs field.

In analogy with the open system dynamics described in Section
\ref{sec:open}, Eq.(\ref{semi}) is equivalent to Eq.(\ref{CTPem})
where the degrees of freedom $x$ of the system are identified with
the metric $g_{\alpha\beta}$ and those of the environment $q$ are
identified with the scalar field $\phi(x)$. However, from the
discussion in the last section it is also clear that Eq.
(\ref{CTPem}) , and hence also the semiclassical Einstein Eq.
(\ref{semi}) results on averaging the full Langevin-type Eq.
(\ref{stochem}) over noise. Thus the semiclassical Einstein equation
incorporates the dissipation but misses out the fluctuation aspect of
the backreaction. The recognition of this crucial point
\cite{HuPhysica} ushered in a new theory known as {\it stochastic
semiclassical gravity}, (or in short, stochastic gravity, as there is
no confusion in this context as to where the stochasticity
originates). Aided by the concept of open systems and the techniques
of the influence functional and the Closed Time Path (CTP) effective
action, stochastic gravity  is the new framework for the
consideration of backreaction because it encompasses fluctuations and
dissipation (from particle creation and other quantum field
processes) on the same footing. Spacetime dynamics is now governed by
a stochastic generalization of the semiclassical Einstein equation
known as the Einstein-Langevin equation,  the analog of Eq.
(\ref{langevin}) in the context of semiclassical gravity (SCG).
Schematically the Einstein-Langevin equation takes on the form
\begin{equation}
   \tilde G_{\mu\nu}(x)
        = \k  \left( T_{\mu\nu}^{\rm c} +  T_{\mu\nu}^{\rm qs}
\right), \quad
   T_{\mu\nu}^{\rm qs}
        \equiv \langle \hat{T}_{\mu\nu} \rangle_{\rm q} +
   T_{\mu\nu}^{\rm s} \label{eq:effective stress tensor}
\end{equation}
Here, $T_{\mu\nu}^c$ is due to classical matter or fields,
$\langle \hat{T}_{\mu\nu} \rangle_q$ is the expectation value of
the stress tensor of the quantum field, and  $T_{\mu\nu}^{\rm qs}$
is a new stochastic term which is related to the quantum
fluctuations of $T_{\mu\nu}$ for the state of the field under
consideration. Taking the average of (\ref{eq:effective stress
tensor}) with respect to the noise distribution will lead to the
conventional semiclassical Einstein equation. It is in this
context that SCG is regarded as a mean field theory.

The fundamentals of this new theory were developed via two
approaches: the axiomatic and the functional. The axiomatic
approach is useful to see the structure of the theory from the
framework of semiclassical gravity, showing the link from the
mean value of the stress-energy tensor to its correlation
functions. The functional approach uses the Feynman-Vernon
influence functional and the Schwinger-Keldysh closed-time-path
effective action methods which are convenient for computations.
It also brings out the open systems concepts and the statistical
and stochastic contents of the theory such as dissipation,
fluctuations, noise and decoherence. There was also theoretical
work on the properties of the stress energy bi-tensor and its
vacuum expectation value, the noise kernel. See, e.g.,
\cite{MV0,MV1,MV2,PH97,HP0,PH1,PH2}. For a broader exploration of
ideas and issues based on this theory read the reviews
\cite{stogra,LivRev}; For a pedagogical introduction with
applications, see \cite{CQGrev,HVErice}.

Thus with the aid of the open system viewpoint it is easy to see
that stochastic gravity is a natural extension of the
well-established semiclassical gravity theory and a useful
framework for the considerations of fluctuations in quantum
matter fields and dissipative dynamics of classical spacetimes,
including metric fluctuations. Stochastic gravity can address
many important issues related to nonequilibrium quantum field
processes in curved spacetimes and find applications to many
problems in gravitation and cosmology.

\subsection{Stochastic Gravity in relation to Quantum Gravity}

Before embarking on the discussion of some applications of stochastic
gravity, let us illustrate the theory with a simple toy model which
minimizes the technical complications. The model will be useful to
clarify the role of the noise kernel and illustrate the relationship
between the semiclassical, stochastic and quantum descriptions.  Let
us assume that the gravitational equations are described by a linear
field $h(x)$ whose source is a massless scalar field $\phi(x)$ which
satisfies the Klein-Gordon equation in flat spacetime $\Box
\phi(x)=0$. The field stress-energy tensor is quadratic in the field,
and independent of $h(x)$. The classical gravitational field
equations will be given by\footnote{In this article we use the
$(+,+,+)$ sign conventions of Ref. \cite{MTW}, and units in which
$c=\hbar=1$.}\begin{equation} \Box h(x)=\kappa T(x), \label{2.12a}
\end{equation}
where $T(x)$ is the (scalar) trace of the stress-energy tensor,
$T(x)=\partial_a\phi(x)\partial^a\phi(x)$ and $\kappa \equiv 16
\pi G$, where G is Newton's constant. Note that this is not a
self-consistent theory since $\phi(x)$ does not react to the
gravitational field $h(x)$. We should also emphasize that this
model is not the standard linearized theory of gravity in which
$T$ is also linear in $h(x)$. It captures, however, some of the
key features of linearized gravity.

In the Heisenberg representation the quantum field $\hat h(x)$
satisfies
\begin{equation}
\Box \hat h(x)=\kappa\hat T(x). \label{2.12}
\end{equation}
Since $\hat T(x)$ is quadratic in the field operator $\hat\phi
(x)$ some regularization procedure has to be assumed in order for
(\ref{2.12}) to make sense. Since we work in flat spacetime we may
simply use a normal ordering prescription to regularize the
operator $\hat T(x)$. The solutions of this equation, i.e. the
field operator at the point $x$, $\hat h(x)$,  may be written in
terms of the retarded propagator $G(x,y)$ as, \begin{equation}
\hat h(x)=\hat h^{(0)}(x) + \kappa \int dx' G(x,x^\prime)\hat
T(x^\prime), \label{2.13}
\end{equation}
where $\hat h^{(0)}(x)$ is the free field which carries information
on the initial conditions and the state of the field. {}From this
solution we may compute, for instance, the symmetric two point
quantum correlation function (the anticommutator)
\begin{equation}
{1\over2}\langle \{\hat h(x),\hat h(y)\}\rangle = {1\over2}
\langle \{\hat h^{(0)}(x),\hat h^{(0)}(y)\}\rangle + {\kappa^2 \over 2}
\int\int dx'dy'G(x,x^\prime)G(y,y^\prime) \langle\{\hat
T(x^\prime),\hat T(y^\prime)\}\rangle, \label{2.14}
\end{equation}
where the expectation value is taken with respect to the quantum
state in which both fields $\phi(x)$ and $h(x)$ are quantized. (We
assume for the free field, $\langle \hat h^{(0)}\rangle=0$.)

We can now consider the semiclassical theory for this problem. If
we assume that $h(x)$ is classical and the matter field is
quantum the semiclassical theory may just be described by
substituting into the classical equation (\ref{2.12a}) the
stress-energy trace by the expectation value of the stress-energy
trace operator $\langle \hat T(x)\rangle$, in some quantum state
of the field $\hat \phi(x)$.  Since in our model $\hat T(x)$ is
independent of $h(x)$ we may simply renormalize its expectation
value using normal ordering, then for the vacuum state of the
field $\hat\phi(x)$, we would simply have $\langle\hat
T(x)\rangle_0=0 $. The semiclassical theory thus reduces to
\begin{equation}
\Box  h(x)=\kappa \langle \hat T(x)\rangle. \label{2.15a}
\end{equation}
The two point function $h(x)h(y)$ that one may derive from this
equation depends on the two point function $\langle \hat
T(x)\rangle \langle \hat T(y)\rangle $ and clearly cannot
reproduce the quantum result (\ref{2.14}) which depends on the
expectation value of the two point operator $\langle\{\hat T(x),\hat
T(y)\}\rangle$. That is, the semiclassical theory entirely misses
the fluctuations of the stress-energy operator $\hat T(x)$.

Let us now see how we can extend the semiclassical theory in order
to account for such fluctuations. The first step is to
characterize these fluctuations. For this, we introduce the noise
kernel as the physical observable that measures the fluctuations
of the stress-energy operator $\hat T$. Define
\begin{equation}
N(x,y)= \frac{1}{2}\langle\{\hat t(x),\hat t(y)\}\rangle
\label{2.14a}
\end{equation}
where $\hat t(x)=\hat T(x)-\langle\hat T(x)\rangle$. The
bi-scalar $N(x,y)$ is real and positive-semidefinite, a
consequence of $\hat t$ being self-adjoint. A simple proof can be
given as follows. Let $|\psi\rangle$ be a given quantum state and
let $\hat Q$ be a self-adjoint operator, $\hat Q^\dagger=\hat Q$,
then one can write $\langle\psi|\hat Q\hat Q|\psi\rangle=
\langle\psi|\hat Q^\dagger Q|\psi\rangle= | \hat
Q|\psi\rangle|^2\geq 0$. Now let $\hat t(x)$ be a self-adjoint
operator, then if we define $\hat Q=\int dx f(x) \hat t(x)$ for
an arbitrary well behaved function $f(x)$, the previous inequality
can be written as $\int dx dy f(x)\langle\psi|\hat t(x)\hat
t(y)|\psi\rangle f(y)\geq 0$, which is the condition for the
noise kernel to be positive semi-definite. Note that when
considering the inverse kernel $N^{-1}(x,y)$, it is implicitly
assumed that one is working in the subspace obtained from the
eigenvectors which have strictly positive eigenvalues when the
noise kernel is diagonalized.

By the positive semi-definite property of the noise kernel
$N(x,y)$ it is possible to introduce a Gaussian stochastic field
as follows:
\begin{equation}
\langle\xi(x)\rangle_s=0,\quad
\langle\xi(x)\xi(y)\rangle_s=N(x,y). \label{2.14b}
\end{equation}
where the subscript $s$ means a statistical average. These
equations entirely define the stochastic process $\xi(x)$ since we
have assumed that it is Gaussian. Of course, higher correlations
could also be introduced but we just try to capture the
fluctuations to lowest order.

The extension of the semiclassical equation may be simply
performed by adding to the right-hand side of the semiclassical
equation (\ref{2.15a}) this stochastic source $\xi(x)$ which
accounts for the fluctuations of $\hat T$ as follows,
\begin{equation}
\Box  h(x)=\kappa\left( \langle \hat T(x)\rangle+\xi(x)\right).
\label{2.15}
\end{equation}
This equation is in the form of a Langevin equation: the field $h(x)$
is classical but stochastic and the observables we may obtain from it
are correlation functions for $h(x)$. In fact, the solution of this
equation may be written in terms of the retarded propagator as,
\begin{equation} h(x)=h^{(0)}(x)+\kappa \int dx^\prime
G(x,x^\prime)\left(\langle\hat T(x^\prime)\rangle
+\xi(x^\prime)\right) , \label{2.16}
\end{equation}
from where the two point correlation function for the classical field
$h(x)$, after using the definition of $\xi(x)$ and that $\langle
h^{(0)}(x)\rangle_s=0$, is given by
\begin{equation}
\langle  h(x) h(y)\rangle_s = \langle  h^{(0)}(x) h^{(0)}(y)\rangle_s
+{\kappa^2 \over 2} \int\int dx^\prime dy^\prime
G(x,x^\prime)G(y,y^\prime) \langle\{\hat T(x^\prime),\hat
T(y^\prime)\}\rangle. \label{2.17}
\end{equation}
Note that in writing $\left<\dots\right>_s$ here we are assuming a
double stochastic average, one is related to the stochastic field
$\xi(x)$ and the other is related to the free field $h^{(0)}(x)$
which is assumed also to be stochastic with a distribution function
to be specified.

Comparing (\ref{2.14}) with (\ref{2.17}) we see that the respective
second term on the right-hand side are identical provided the
expectation values are computed in the same quantum state for the
field $\hat \phi(x)$ (recall that we have assumed $T(x)$ does not
depend on $h(x)$). The fact that the field $h(x)$ is also quantized
in (\ref{2.14}) does not change the previous statement. The nature of
the first term on the right-hand side of Eqs. (\ref{2.14}) and
(\ref{2.17}) is different: in the first case it is the two point
quantum expectation value of the free quantum field $\hat h^{(0)}$
whereas in the second case it is the stochastic average of the two
point classical homogeneous field $h^{(0)}$, which depends on the
initial conditions. Now we can still make these terms equal to each
other if we assume for the homogeneous field $h^{(0)}$ a Gaussian
distribution of initial conditions such that
\begin{equation}
\langle  h^{(0)}(x) h^{(0)}(y)\rangle_s=  \frac{1}{2}\langle\{\hat
h^{(0)}(x),\hat h^{(0)}(y)\}\rangle. \label{2.17a}
\end{equation}
This Gaussian stochastic field $h^{(0)}(x)$ can always be defined due
to the positivity of the anti-commutator. Thus, under this assumption
on the initial conditions for the field $h(x)$ the two point
correlation function of (\ref{2.17}) equals the quantum expectation
value of (\ref{2.14}) exactly. An interesting feature of the
stochastic description is that the quantum anticommutator of
(\ref{2.14}) can be written as the right-hand side of equation
(\ref{2.17}), where the first term contains all the information on
the initial conditions for the stochastic field $h(x)$ and the second
term codifies all the information on the quantum correlations of the
source. This separation is also seen in the description of some
quantum Brownian motion models which are typically used as paradigms
of open quantum systems \cite{CRVopensys,CRVtunnel}.

It is interesting to note that in the standard linearized theory
of gravity $T(x)$ depends also on $h(x)$, both explicitly and also
implicitly through the coupling of $\phi(x)$ with $h(x)$. The
equations are not so simple but it is still true that the
corresponding Langevin equation leads to the correct symmetrized
two point quantum correlations for the metric perturbations
\cite{MV2,RouVer03b}. Thus in a linear theory as in the model just
described one may just use the statistical description given by
(\ref{2.15}) to compute the symmetric quantum two point function
of equation (\ref{2.13}). This does not mean that we can recover
all quantum correlation functions with the stochastic
description, see Ref. \cite{CRVopensys} for a general discussion
about this point. Note that, for instance, the commutator of the
classical stochastic field $h(x)$ is obviously zero, but the
commutator of the quantum field $\hat h(x)$ is not zero for
timelike separated points; this is the prize we pay for the
introduction of the classical field $\xi(x)$ to describe the
quantum fluctuations. Furthermore, the statistical description is
not able to account for the graviton-graviton effects which go
beyond the linear approximation in $\hat h(x)$.

\section{Black Hole Fluctuations and Backreaction}

As the first example we consider the backreaction of Hawking
radiation on black holes \cite{bhbkr} with fluctuations, i.e., how a
quantum field and its fluctuations influence the behavior of the
background spacetime. We will only sketch the strategy of this
program based on stochastic gravity, as detailed calculations are
still in progress. The formalism described in Section \ref{sec:open}
will be useful. Here we study the simpler class of problems of a
quasi-static black hole in quasi-equilibrium (a box is required) with
its Hawking radiation described by a scalar field. The goal is to
obtain a stochastic influence action analogous to (\ref{Seffxi}) for
this model of a black hole coupled to a scalar field. From it one can
derive an Einstein-Langevin equation analogous to (\ref{langevin}).

We consider the simplest model of this class described by a
perturbed Schwarzschild metric, used by York \cite{York} to
analyze black hole backreaction. We focus on the new aspects of
noise and fluctuations, their origin and  attributes.


In this model the black hole spacetime is described by a
spherically symmetric static nonvacuous metric with line element of the
following general form written in advanced time
Eddington-Finkelstein coordinates \be ds^2 =
g_{\mu\nu}dx^{\mu}dx^{\nu} = -e^{2\psi}\left(1 - {2m\over
r}\right)dv^2 + 2 e^{2\psi}dvdr + r^2~d{\Omega}^2
\label{ssmetric} \te where $\psi = \psi(r)$ and $m = m(r)$ , $ v
= t + r + 2Mln\left({r\over 2M} -1 \right)$ and $d{\Omega}^2$ is
the line element on the two sphere. Hawking radiation is
described by a massless, conformally coupled quantum scalar field
$\phi$ with the classical action \be S_m[\phi, g_{\mu\nu}] =
-{\ha}\int d^n x \sqrt{-g}[g^{\mu\nu}\partial_{\mu}\phi
\partial_{\nu}\phi + \xi(n) R{\phi}^2] \label{phiact} \te where
$\xi(n) = {(n-2)\over 4(n-1)}$ ($n$ is the dimension of
spacetime) and $R$ is the curvature scalar of the spacetime it
lives in.


Let us consider linear perturbations $h_{\mu\nu}= g_{\mu\nu} -
g^{(0)}_{\mu\nu}$ off a background Schwarzschild metric
$g^{(0)}_{\mu\nu}$ with line element \be  (ds^2)^0 = \left( 1 -
{2M\over r}\right)dv^2 + 2dvdr + r^2d{\Omega}^2 \label{schwarz} \te
We look for this class of perturbed metrics in the form given by
(\ref{ssmetric}), (thus restricting our consideration only to
spherically symmetric perturbations): \be e^\psi \simeq  1+ \epsilon
\rho(r),   \ \  m \simeq M[ 1 + \epsilon \mu (r)] \label{mu} \te
where ${\epsilon\over \lambda M^2} = {1\over 3}a T_H^4 ; a
={{\pi}^2\over 30} ; \lambda = 90(8^4)\pi^2$. $T_H$ is the Hawking
temperature. (This particular parametrization of the perturbation is
chosen following York's \cite{York} notation.) Thus the only non-zero
components of $h_{\mu\nu}$ are \be h_{vv} = -\left((1 - {2M\over
r})2\epsilon \rho(r) + {2M\epsilon \mu (r)\over r}\right), \ \ h_{vr}
= \epsilon\rho (r) \label{hvr} \te So this represents a metric with
small static and radial perturbations about a Schwarzschild black
hole. The initial quantum state of the scalar field is taken to be the
Hartle-Hawking vacuum, which is essentially a thermal state at the
Hawking temperature as far as static observers are concerned
\footnote{The Hartle-Hawking vacuum is a pure state which is
perceived as vacuum by free-falling observers crossing either the
black hole or the white hole horizons. However, if one traces out the
modes localized in the second asymptotically flat region, one ends up
with an incoherent density matrix perceived as a thermal state by
static observers in the first asymptotically flat region.} and it
represents a black hole in (unstable) thermal equilibrium with its own
Hawking radiation.

The metric perturbation expansion induces a decomposition of the
Einstein tensor  $G_{\mu\nu} \simeq G^{(0)}_{\mu\nu} + \delta
G_{\mu\nu}$ where $G^{(0)}_{\mu\nu}$ is the Einstein tensor for the
background spacetime. The zeroth order solution gives a background
metric in empty space, i.e, the Schwarzschild metric. $\delta
G_{\mu\nu}$ is the linear correction to the Einstein tensor in the
perturbed metric. The semiclassical Einstein equation  in this
approximation therefore reduces to $$ \delta G_{\mu\nu}(g^{(0)}, h) =
\kappa \langle T_{\mu\nu}\rangle $$ York solved this equation to
first order by using the expectation value of the energy momentum
tensor for a conformally coupled scalar field in the Hartle-Hawking
vacuum in the unperturbed (Schwarzschild) spacetime on the right hand
side and $\delta G_{\mu\nu}$ on the left hand side is calculated
using (\ref{hvr}). This yields the corrections to the background
metric induced by the backreaction encoded in the functions $\mu(r)$
and $\rho(r)$, which amounts to the noise-averaged backreaction
effects. We are interested in the fluctuations and its effects.


We now derive the CTP effective action for this model, following
the treatment of Ref. \cite{CamHu}. Using the metric
(\ref{schwarz}) (and neglecting the surface terms that appear in
an integration by parts) we have  the action for the scalar
field  written perturbatively as
\begin{equation}
   S_m[\phi,h_{\mu\nu}]
        \ = \  {1\over 2}\int d^nx{\sqrt{-g^{(0)}}}\ \phi
               \left[ \Box + V^{(1)} + V^{(2)} + \cdots
              \right] \phi,
\label{phipert}
\end{equation}
where the first and second order perturbative operators $V^{(1)}$
and $V^{(2)}$ are given by
\begin{eqnarray}
V^{(1)}   & \ \equiv \ & - {1\over \sqrt{-g^{(0)}}} \left\{
[\partial_\mu\left(\sqrt{-g^{(0)}}\bar h^{\mu\nu}(x)\right)]
                                \partial_\nu
                              +\bar h^{\mu\nu}(x)\partial_\mu
                              \partial_\nu
                            +\xi(n) R^{(1)}(x)
                     \right\},
               \nonumber \\
V^{(2)}
    & \ \equiv \ & - {1\over \sqrt{-g^{(0)}}}
 \left\{ [\partial_\mu \left(\sqrt{-g^{(0)}} \hat h^{\mu\nu}(x)\right)] \partial_\nu
 +\hat h^{\mu\nu}(x)\partial_\mu \partial_\nu
    - \xi(n) \left( R^{(2)}(x) +{1\over 2}h(x)R^{(1)}(x) \right) \right\}.
\end{eqnarray}
In the above expressions, $R^{(k)}$ is the $k$-order term in the
pertubation $h_{\mu\nu}(x)$ of the scalar curvature $R$ and $\bar
h_{\mu\nu}$ and $\hat h_{\mu\nu}$ denote a linear and a quadratic
combination of the perturbation, respectively,
\begin{equation}
   \bar h_{\mu\nu}
          \equiv   h_{\mu\nu} - {1\over 2} h g^{(0)}_{\mu\nu};
          \;\;
   \hat h_{\mu\nu}
          \equiv   h^{\,\, \alpha}_\mu h_{\alpha\nu}
                      -{1\over 2} h h_{\mu\nu}
                      +{1\over 8} h^2 g^{(0)}_{\mu\nu}
                      -{1\over 4} h_{\alpha\beta}h^{\alpha\beta}
g^{(0)}_{\mu\nu}.
   \label{eq:def bar h}
\end{equation}
 From quantum field theory in curved spacetime considerations we
take the following action for the gravitational field (see
\cite{MV1,CamHu} for more details)
\begin{equation}
   S^{(div)}_g[g_{\mu\nu}]
         \ = \ {1\over\ell^{n-2}_P}\int d^nx\ \sqrt{-g}R(x)
                +{\alpha\bar\mu^{n-4}\over4(n-4)}
                   \int d^nx\ \sqrt{-g}
                   \left[ 3R_{\mu\nu\alpha\beta}(x)
                           R^{\mu\nu\alpha\beta}(x)
                    \left( 1-360(\xi(n) - {1\over6})^2
                          \right)R(x)R(x)          \right].
\end{equation}
The first term is the classical Einstein-Hilbert action and the
second term is the counterterm in four dimensions used  to
renormalize the divergent effective action. In this action
$\ell^2_P = 16\pi G$, $\alpha = (2880\pi^2)^{-1}$ and $\bar\mu$
is the renormalization mass scale.

We are interested in computing the CTP effective action for the model
given by the form (\ref{phipert}) for the matter action and when the
field $\phi$ is initially in the Hartle- Hawking (HH) vacuum. Since
the initial state of the field is described by a thermal density
matrix at the HH temperature $T_H$\footnote{As mentioned earlier,
this is true for the basis of modes associated to static observers
and provided that one is not concerned about quantum correlations
(entanglement) between the two asymptotically flat regions, so that
the modes localized in the second region can be traced out. Otherwise
the full Hartle-Hawking vacuum, which is a pure state, should be
considered.},
the finite temperature CTP effective action ($T \equiv 1/\beta$) for
this model is given by (for details see \cite{CamHu})
\begin{equation}
   \Gamma^\beta_{CTP}[h^\pm_{\mu\nu}]
        \ = \ S^{(div)}_g[h^+_{\mu\nu}]
             -S^{(div)}_g[h^-_{\mu\nu}]
             -{i\over2}Tr\{ \ln\bar G^\beta_{ab}[h^\pm_{\mu\nu}]\},
   \label{eq:eff act two fields}
\end{equation}
where $\pm$ denote the forward and backward time path of the CTP
formalism and $\bar G^\beta_{ab}[h^\pm_{\mu\nu}]$ is the complete
$2\times 2$ matrix propagator ($a$ and $b$ take $\pm$ values:
$G_{++},G_{+-}$ and $G_{--}$ correspond to the Feynman, Wightman and
Schwinger Green's functions respectively) with thermal boundary
conditions for the differential operator $\sqrt{-g^{(0)}}(\Box +
V^{(1)} + V^{(2)} + \cdots)$. The actual form of $\bar G^\beta_{ab}$
cannot be explicitly given. However, it is easy to obtain a
perturbative expansion in terms of $V^{(k)}_{ab}$, the $k$-order
matrix version of the complete differential operator defined by
$V^{(k)}_{\pm\pm} \equiv \pm V^{(k)}_{\pm}$ and $V^{(k)}_{\pm\mp}
\equiv 0$, and $G^\beta_{ab}$, the thermal matrix propagator for a
massless scalar field in Schwarzschild spacetime . To second order
$\bar G^\beta_{ab}$ reads
\begin{eqnarray}
   \bar G^\beta_{ab}
        \ = \  G^\beta_{ab}
              -G^\beta_{ac}V^{(1)}_{cd}G^\beta_{db}
              -G^\beta_{ac}V^{(2)}_{cd}G^\beta_{db}
              +G^\beta_{ac}V^{(1)}_{cd}G^\beta_{de}
               V^{(1)}_{ef}G^\beta_{fb}
              +\cdots
\end{eqnarray}
Expanding the logarithm and dropping one term independent of the
perturbation $h^\pm_{\mu\nu}(x)$, the CTP effective action may be
perturbatively written as
\begin{eqnarray}
   \Gamma^\beta_{CTP}[h^\pm_{\mu\nu}]
        & \ = \ &  S^{div}_g[h^+_{\mu\nu}] - S^{div}_g[h^-_{\mu\nu}]
                +{i\over2}Tr[ V^{(1)}_{+}G^\beta_{++}
                               -V^{(1)}_{-}G^\beta_{--}
                               +V^{(2)}_{+}G^\beta_{++}
                               -V^{(2)}_{-}G^\beta_{--}
                              ]
                \nonumber \\
        &       & -{i\over4}Tr[  V^{(1)}_{+}G^\beta_{++}
                                 V^{(1)}_{+}G^\beta_{++}
                               + V^{(1)}_{-}G^\beta_{--}
                                 V^{(1)}_{-}G^\beta_{--}
                               -2V^{(1)}_{+}G^\beta_{+-}
                                 V^{(1)}_{-}G^\beta_{-+}
                              ].
   \label{eq:effective action}
\end{eqnarray}
However, unlike the case of Ref.\cite{CamHu} where $h_{\mu\nu}$
represented a perturbation about flat space and hence one had
knowledge of exact ``unperturbed" thermal propagators, in this case,
since the perturbation is about Schwarzschild spacetime, exact
expressions for the corresponding unperturbed propagators
$G^\beta_{ab}[h^\pm_{\mu\nu}]$ are not known. Therefore apart from
the approximation of computing the CTP effective action to certain
order in perturbation theory, an appropriate approximation scheme for
the unperturbed Green's functions is also required. York used the
Page approximation \cite{Page82} for $\langle T_{\mu \nu}\rangle$ in
the Schwarzschild metric.  The additional complication here is that
while to obtain $\langle T_{\mu\nu}\rangle$ the knowledge of only the
thermal Feynman Green's function is required, to calculate the CTP
effective action one needs the knowledge of the full matrix
propagator, which involves the Feynman, Schwinger and Wightman
functions. We can put aside the technical complexity in the
calculation of the full thermal matrix propagator
$G^\beta_{ab}[h^\pm_{\mu\nu}]$ as our main interest is to identify
and analyze the noise term which is the new ingredient in the
backreaction problem. We have mentioned that the noise term gives a
stochastic contribution $T_{\mu\nu}^{\rm s}$ to the Einstein Langevin
equation (\ref{eq:effective stress tensor}). We have also stated that
this term is related to the variance of fluctuations in $T_{\mu\nu}$,
i.e., schematically, to $\langle T^2_{\mu\nu} \rangle$. Since the
Influence Functional or CTP formalism itself does not depend on the
nature of the approximation, we will attempt to exhibit the general
structure and project what is to be expected.

If we denote the difference and the sum of the perturbations
$h^\pm_{\mu\nu}$  by $[h_{\mu\nu}] \equiv h^+_{\mu\nu} -
h^-_{\mu\nu}$ and $\{h_{\mu\nu}\} \equiv h^+_{\mu\nu} +
h^-_{\mu\nu}$, respectively, the influence functional form of the
thermal CTP effective action may be written to second order in
$h_{\mu\nu}$ as \cite{CamHu}
\begin{eqnarray}
   \Gamma^\beta_{CTP}[h^\pm_{\mu\nu}]
         & \simeq &  {1\over2\ell^2_P}\int d^4x\ d^4x'\
                       [h_{\mu\nu}](x){L}_{(o)}^{\mu\nu,\alpha\beta}(x,x')
                       \{h_{\alpha\beta}\}(x')
                    +{1\over2}\int d^4x\
                       [h_{\mu\nu}](x)T^{\mu\nu}_{(\beta)}(x)
                   \nonumber \\
        &            & +{1\over2}\int d^4x\ d^4x'\
                       [h_{\mu\nu}](x){H}^{\mu\nu,\alpha\beta}(x,x')
                       \{h_{\alpha\beta}\}(x')
                     -{1\over2}\int d^4x\ d^4x'\
                       [h_{\mu\nu}](x)D^{\mu\nu,\alpha\beta}(x,x')
                       \{h_{\alpha\beta}\}(x')
                     \nonumber \\
        &            &+{i\over2}\int d^4x\ d^4x'\
                       [h_{\mu\nu}](x)N^{\mu\nu,\alpha\beta}(x,x')
                      [h_{\alpha\beta}](x').
\label{CTPbh}
\end{eqnarray}
The first term is the Einstein-Hilbert action to second order in the
perturbation $h^\pm_{\mu\nu}(x)$ and $\kl{L}{(o)}{x}$ is a symmetric
local kernel, {\sl i.e.} $\kl{L}{(o)}{x,x'}$ =
$\kl{L}{(o)}{x',x}$. The second is a local term linear in
$h^\pm_{\mu\nu}(x)$.  $T_{(\beta)}^{\mu\nu}(x)$ represents the zeroth
order contribution to $\langle \hat{T}_{\mu\nu}(x)\rangle$ and far
away from the horizon it takes the form of the stress tensor of
massless scalar particles at temperature $\beta^{-1}$. The third and
fourth terms constitute the remaining quadratic component of the real
part of the effective action. The kernels
$H^{\mu\nu,\alpha\beta}(x,x')$ and $D^{\mu\nu,\alpha\beta}(x,x')$ are
respectively even and odd in $x,x'$. The last term gives the imaginary
part of the effective action and the kernel
$N^{\mu\nu,\alpha\beta}(x,x')$ is symmetric.  This is the general
structure of the CTP effective action arising from the calculation of
the traces in equation (\ref{eq:effective action}). Of course, to
write down explicit expressions for the non-local kernels one requires
the input of the explicit form of $G^\beta_{ab}[h^\pm_{\mu\nu}]$ ,
which we have not used. In spite of this limitation we can make some
interesting observations from this effective action. Connecting this
thermal CTP effective action to the influence functional via equation
(\ref{CTPIFconn}) we see that the nonlocal imaginary term containing
the kernel $N^{\mu\nu,\alpha\beta}(x,x')$ is responsible for the
generation of the stochastic noise term in the Einstein-Langevin
equation and the real non-local term containing kernel
$D^{\mu\nu,\alpha\beta}(x,x')$ is responsible for the non-local
dissipation term. The Einstein-Langevin equation can be generated from
equation (\ref{stochem}) by first constructing the improved
semiclassical effective action in accordance with (\ref{Seffxi}) and
deriving the equation of motion (\ref{stochem}) by taking a functional
derivative of the above effective action with respect to
$[h_{\mu\nu}]$ and equating it to zero. With the identification of the
noise and dissipation kernels, one can use them to write down a
Fluctuation-Dissipation relation (FDR) analogous to (\ref{FDR}) in the
context of black holes.

\section{Structure Formation in Inflationary Universes}
\label{sec:strfor}

Cosmological structure formation is a key problem in modern
cosmology \cite{Pad93,Smoot92} and inflation
\cite{infcos,KolTur90,Linde90} offers a natural solution to this
problem. If an inflationary period is present, the initial seeds for
the generation of the primordial inhomogeneities that lead to the
large scale structure observed in the present universe have their
source in the quantum fluctuations of the inflaton field, the field
which is generally responsible for driving inflation. Stochastic
gravity provides a sound and natural formalism for the derivation of
the cosmological perturbations generated during inflation.

In Ref.~\cite{RouVer03a} it was shown that the correlation
functions that follow from the Einstein-Langevin equation which
emerges in the framework of stochastic gravity coincide with that
obtained with the usual quantization procedures \cite{MukFelBra92}
when both the metric perturbations and the inflaton fluctuations
are both linearized. Stochastic gravity, however, can naturally
deal with the fluctuations of the inflaton field even beyond the
linear approximation.

Here we will illustrate the equivalence with the usual formalism,
based on the quantization of the linear cosmological and inflaton
perturbations, with one of the simplest chaotic inflationary
models in which the background spacetime is a quasi de Sitter
universe \cite{RouVer00,RouVer03a}.




In this chaotic inflationary model \cite{infcos,Linde90} the inflaton
field $\phi$ of mass $m$ is described by the following Lagrangian
density
\begin{equation} {\cal
L}(\phi)={1\over 2}g^{ab}\nabla_a\phi \nabla_b\phi + {1\over
2}m^2\phi^2. \label{1.14}
\end{equation}
The conditions for the existence of an inflationary period, which is
characterized by an accelerated cosmological expansion, is that the
value of the field over a region with the typical size of the Hubble
radius is higher than the Planck mass $m_P$. In order to solve the
cosmological horizon and flatness problem more than 60 e-folds of
expansion are needed; to achieve this the scalar field should begin
with a value higher than $3m_P$. The inflaton mass is small: as we
will see, the large scale anisotropies measured in the cosmic
background radiation restrict the inflaton mass to be of the order of
$10^{-6}m_P$. We will not discuss the naturalness of this inflationary
model and we will simply assume that if one such region is found
(inside a much larger universe) it will inflate to become our
observable universe.

We want to study the metric perturbations produced by the
stress-energy tensor fluctuations of the inflaton field on the
homogeneous background of a flat Friedmann-Robertson-Walker
model, described by the cosmological scale factor $a(\eta)$,
where $\eta$ is the conformal time, which is driven by the
homogeneous inflaton field $\phi(\eta)=\langle\hat\phi\rangle$.
Thus we write the inflaton field in the following form:
$\hat\phi=\phi(\eta)+ \hat\varphi (x)$,
where $\hat\varphi (x)$ corresponds to a free massive quantum
scalar field with zero expectation value on the homogeneous
background metric: $\langle\hat\varphi\rangle=0$. We will
restrict ourselves to scalar-type metric perturbations because
these are the ones that couple to the inflaton fluctuations in
the linear theory. We note that this is not so if we were to
consider inflaton fluctuations beyond the linear approximation,
then tensorial and vectorial metric perturbations would also be
driven. The perturbed metric $\tilde g_{ab}=g_{ab}+h_{ab}$ can be
written in the longitudinal gauge as,
\begin{equation}
ds^2=a^2(\eta)[-(1+2\Phi(x))d\eta^2+(1-2\Psi(x))\delta_{ij}dx^idx^j],
\label{1.16} \end{equation} where the scalar metric perturbations
$\Phi(x)$ and $\Psi(x)$ correspond to Bardeen's gauge invariant
variables \cite{Bar80}.


\subsection{Einstein-Langevin equation for scalar metric perturbations}


The Einstein-Langevin equation 
is gauge invariant, thus we can work in a desired gauge and then
extract the gauge invariant quantities. It is given by
\begin{equation}
G^{(0)}_{ab}-8\pi G\langle\hat T^{(0)}_{ab}\rangle+
G^{(1)}_{ab}(h)-8\pi G\langle\hat T^{(1)}_{ab}(h)\rangle=
 8\pi G\xi_{ab},
\label{1.17}
\end{equation}
where the two first terms cancel, that is $G^{(0)}_{ab}-8\pi
G\langle\hat T^{(0)}_{ab}\rangle=0$, as the background metric
satisfies the semiclassical Einstein equations. Here the
subscripts $(0)$ and $(1)$ refer to functions in the background
metric $g_{ab}$ and linear in the metric perturbation $h_{ab}$,
respectively. The stress tensor operator $\hat T_{ab}$ for the
minimally coupled inflaton field in the perturbed metric is:
\begin{equation}
\hat T_{ab}= \tilde\nabla_{a}\hat\phi
\tilde\nabla_{b}\hat\phi+{1\over2}\tilde g_{ab} (\tilde\nabla_{c}
\hat\phi \tilde\nabla^{c}\hat\phi+ m^2\hat\phi^2). \label{1.18}
\end{equation}

Using the decomposition of the scalar field into its homogeneous
and inhomogeneous part and the metric
$\tilde g_{ab}$ into its homogeneous background $g_{ab}$ and its
perturbation $h_{ab}$, the renormalized expectation value for the
stress-energy tensor operator can be written as
\begin{equation}
\langle \hat T^R_{ab}[\tilde g]\rangle= \langle \hat
T_{ab}[\tilde g]\rangle_{\phi\phi}+ \langle \hat T_{ab}[\tilde
g]\rangle_{\phi\varphi}+ \langle \hat T^R_{ab}[\tilde
g]\rangle_{\varphi\varphi}, \label{1.19}
\end{equation}
where the subindices indicate the degree of dependence on the
homogeneous field $\phi$ and its perturbation $\varphi$. The
first term in this equation depends only on the homogeneous field
and it is given by the classical expression. The second term is
proportional to $\langle\hat\varphi[\tilde g]\rangle$ which is
not zero because the field dynamics is considered on the
perturbed spacetime, {\it i.e.}, this term includes the coupling
of the field with $h_{ab}$ and may be obtained from the
expectation value of the linearized Klein-Gordon equation,
$\left( \Box_{g+h}-m^2\right)\hat\varphi =0. \label{1.19a}$
The last term in Eq.~(\ref{1.19}) corresponds to the expectation
value to the stress tensor for a free scalar field on the
spacetime of the perturbed metric.

After using the previous decomposition, the noise kernel
$N_{abcd}[g;x,y)$ 
can be written as
\begin{equation}
\langle \{\hat t_{ab}[g;x),\hat t_{cd}[g;y)\}\rangle \ = \ \langle
\{\hat t_{ab}[g;x),\hat t_{cd}[g;y)\} \rangle_{(\phi\varphi)^2} +
\langle \{\hat t_{ab}[g;x),\hat t_{cd}[g;y)\}
\rangle_{(\varphi\varphi)^2}, \label{1.20}
\end{equation}
where we have used the fact that $\langle\hat\varphi\rangle=0
=\langle\hat\varphi\hat\varphi\hat\varphi\rangle$ for Gaussian
states on the background geometry. We consider the vacuum state
to be the Euclidean vacuum which is preferred in the de Sitter
background, and this state is Gaussian. In the above equation the
first term is quadratic in $\hat\varphi$ whereas  the second one
is quartic, both contributions to the noise kernel are separately
conserved since both $\phi(\eta)$ and $\hat\varphi$ satisfy the
Klein-Gordon field equations on the background spacetime.
Consequently, the two terms can be considered separately. On the
other hand if one treats $\hat \varphi$ as a small perturbation
the second term in (\ref{1.20}) is of lower order than the first
and may be consistently neglected, this corresponds to neglecting
the last term of Eq.~(\ref{1.19}). The stress tensor fluctuations
due to a term of that kind were considered in Ref.
\cite{RouVer00}.

We can now write down the Einstein-Langevin equations (\ref{1.17}) to
linear order in the inflaton fluctuations. It is easy to check
\cite{RouVer03a} that the {\it space-space} components coming from
the stress tensor expectation value terms and the stochastic tensor
are diagonal, i.e. $\langle\hat T_{ij}\rangle=0= \xi_{ij}$ for
$i\not= j$. This, in turn, implies that the two functions
characterizing the scalar metric perturbations are equal: $\Phi=\Psi$
in agreement with Ref. \cite{MukFelBra92}. The equation for $\Phi$
can be obtained from the $0i$-component of the Einstein-Langevin
equation, which in Fourier space reads
\begin{equation} 2ik_i({\cal H}\Phi_k+\Phi'_k)= 8\pi
G(\xi_{0i})_k, \label{1.21}
\end{equation}
where $k_i$ is the comoving momentum component associated to the
comoving coordinate $x^i$, and we have used the definition
$\Phi_k(\eta)= \int d^3 x \exp(-i\vec k\cdot\vec x)\Phi(\eta,\vec
x)$. Here primes denote derivatives with respect to the conformal
time $\eta$ and ${\cal H}=a'/a$. A nonlocal term of dissipative
character which comes from the second term in Eq.~(\ref{1.19}) should
also appear on the left hand side of Eq.~(\ref{1.21}), but we have
neglected it to simplify the forthcoming expressions\footnote{Such a
term, which leads to an integrodifferential Einstein-Langevin
equation, might not be negligible in some situations. Despite the
apparent difficulty of dealing with an integrodifferntial equation,
in this case the Einstein-Langevin equation can be actually
transformed, after suitable manipulation, into an ordinary
differential equation, so that the inclusion of the nonlocal term is
still tractable \cite{RouVer03a}}. Note, however, that the
equivalence of the stochastic approach to linear order in
$\hat\varphi$ and the usual linear cosmological perturbations
approach is independent of that approximation \cite{RouVer03a}. To
solve Eq.~(\ref{1.21}), whose left-hand side comes from the
linearized Einstein tensor for the perturbed metric
\cite{MukFelBra92}, we need the retarded propagator for the
gravitational potential $\Phi_k$,
\begin{equation} G_k(\eta,\eta')= -i {4\pi\over k_i m_P^2}\left(
\theta(\eta-\eta') {a(\eta')\over a(\eta)}+f(\eta,\eta')\right),
\label{1.22} \end{equation} where $f$ is a homogeneous solution
of Eq.~(\ref{1.21}) related to the initial conditions chosen and
$m_P^2=1/G$. For instance, if we take
$f(\eta,\eta')=-\theta(\eta_0-\eta')a(\eta')/a(\eta)$ the
solution would correspond to ``turning on" the stochastic source
at $\eta_0$. With the solution of the Einstein-Langevin equation
(\ref{1.21}) for the scalar metric perturbations we are in a
position to compute the two-point correlation functions for these
perturbations.


\subsection{Correlation functions for scalar metric perturbations}


The two-point correlation function for the scalar metric
perturbations induced by the inflaton fluctuations is thus given
by
\begin{equation}
\langle\Phi_k(\eta)\Phi_{k'}(\eta')\rangle_s = (2\pi)^2\delta(\vec
k+\vec k') \times\int^\eta \!d\eta_1\int^{\eta'}\!d\eta_2
G_k(\eta,\eta_1) G_{k'}(\eta',\eta_2)
\langle(\xi_{0i})_k(\eta_1)(\xi_{0i})_{k'}(\eta_2)\rangle_s .
\label{1.23}
\end{equation}
Here the two-point correlation function for the stochastic source,
which is connected to the stress-energy tensor fluctuations
through the noise kernel is given by,
\begin{equation}
\langle (\xi_{0i})_k(\eta_1)(\xi_{0i})_{-k}(\eta_2)\rangle_s \ = \
{1\over2} \langle\{(\hat t_{0i})_k(\eta_1),(\hat
t_{0i})_{-k}(\eta_2)\}\rangle_{\phi\varphi} \ = \
{1\over2}k_ik_i\phi'(\eta_1)\phi'(\eta_2)G_k^{(1)}(\eta_1,\eta_2),
\label{1.24}
\end{equation}
where $G_k^{(1)}(\eta_1,\eta_2)=\langle\{\hat\varphi_k(\eta_1),
\hat\varphi_{-k}(\eta_2)\}\rangle$ is the $k$-mode Hadamard
function for a free minimally coupled scalar field in the
appropriate vacuum state on the Friedmann-Robertson-Walker
background.

In practice, to make the explicit computation of the Hadamard
function we will assume that the field state is in the Euclidean
vacuum and the background spacetime is de Sitter. Furthermore, we
will compute the Hadamard function for a massless field, and will
make a perturbative expansion in terms of the dimensionless
parameter $m/m_P$. Thus we consider
$$\bar
G_k^{(1)}(\eta_1,\eta_2) = \langle 0|\{\hat y_k(\eta_1),\hat
y_{-k}(\eta_2)\}|0\rangle = 2{\cal
R}\left(u_k(\eta_1)u_k^*(\eta_2)\right),$$ with $\hat y_k(\eta)=
a(\eta)\hat\varphi_k(\eta)= \hat a_k u_k(\eta)+\hat
a_{-k}^\dagger u_{-k}^*(\eta)$ and where
$u_k=(2k)^{-1/2}e^{ik\eta}(1-i/\eta)$ are the positive frequency
$k$-mode for a massless minimally coupled scalar field on a de
Sitter background, which define the Euclidean vacuum state: $\hat
a_k|0\rangle=0$ \cite{BirDav}.

The assumption of a massless field for the computation of the
Hadamard function is made because massless modes in de Sitter are
much simpler to deal with than massive modes. We can see that this
is, however, a reasonable approximation as follows. For a given
mode the $m=0$ approximation is reasonable when its wavelength
$\lambda$ is shorter that the Compton wavelength, $\lambda_c=1/m$.
In our case we have a very small mass $m$ and the horizon size
$H^{-1}$, where $H$ is the Hubble constant $H=\dot a/a$ (here
$a(t)$ with $t$ the physical time $dt=ad\eta$) satisfies that
$H^{-1}<\lambda_c$. Thus, for modes inside the horizon
$\lambda<\lambda_c$ and $m=0$ is a reasonable approximation.
Outside the horizon massive modes decay in amplitude as $\sim \exp
(-m^2 t/H)$ whereas massless modes remain constant, thus when
modes leave the horizon the approximation will eventually break
down. However, we only need to ensure that the approximation is
still valid after $60$ e-folds, {\it i.e.} $Ht\sim 60$, but this
is the case since $60\; m^2< H^2$ given that $m\sim 10^{-6}m_P$,
and $m\ll H$ as in most inflationary models \cite{KolTur90,Pad93}.

The background geometry is not exactly that of de Sitter
spacetime, for which $a(\eta)=-(H\eta)^{-1}$ with $-\infty <\eta<
0$. One can expand in terms of the ``slow-roll" parameters and
assume that to first order $\dot\phi(t)\simeq m_P^2(m/m_P)$, where
$t$ is the physical time. The correlation function for the metric
perturbation (\ref{1.23}) can then be easily computed; see
Ref.~\cite{RouVer00,RouVer03a} for details. The final result,
however, is very weakly dependent on the initial conditions as one
may understand from the fact that the accelerated expansion of de
quasi-de Sitter spacetime during inflation erases the information
about the initial conditions. Thus one may take the initial time
to be $\eta_0=-\infty$ and obtain to lowest order in $m/m_P$ the
expression
\begin{equation}
\langle\Phi_k(\eta)\Phi_{k'}(\eta')\rangle_s\simeq 8\pi^2 (m/m_P)^2
k^{-3}(2\pi)^3\delta(\vec k+\vec k') \cos k(\eta-\eta'). \label{1.25}
\end{equation}
\indent From this result two main conclusions are derived. First, the
prediction of an almost Harrison-Zel'dovich scale-invariant spectrum
for large scales, i.e. small values of $k$. Second, since the
correlation function is of order of $(m/m_P)^2$ a severe bound to the
mass $m$ is imposed by the gravitational fluctuations derived from
the small values of the Cosmic Microwave Background (CMB)
anisotropies detected by COBE \cite{MukFelBra92}. This bound is of
the order of $(m/m_P)\sim 10^{-6}$ .

We should now comment on some differences with those works in
Ref.~\cite{strforRenegades} which used a self-interacting scalar
field or a scalar field interacting nonlinearly with other
fields. In those works an important relaxation of the ratio
$m/m_p$ was found. The long wavelength modes of the inflaton
field  were regarded as an open system in an environment made out
of the shorter wavelength modes. Then, Langevin type equations
were used to compute the correlations of the long wavelength
modes driven by the fluctuations of the shorter wavelength modes.
In order to get a significant relaxation on the above ratio,
however, one had to assume that the correlations of the free long
wavelength modes, which correspond to the dispersion of the
system initial state, had to be very small. Otherwise they
dominate by several orders of magnitude those fluctuations that
come from the noise of the environment. This would require a
great amount of fine-tuning for the initial quantum state of each
mode \cite{RouVer03a}. We should remark that in the model
discussed here there is no environment for the inflaton
fluctuations. The inflaton fluctuations, however, are responsible
for the noise that induce the metric perturbations.\\

{\bf Acknowledgements} This work is supported in part by NSF grant
PHY98-00967, the MICYT Research Project No. FPA-2001-3598 and
European project HPRN-CT-2000-00131.


\begin{thebibliography}{99}
\bibitem{HMLA}  B. L. Hu and A. Matacz, ``Quantum Noise in Gravitation and
Cosmology'' in {\it Fluctuations and Order}, edited by Marko
Millonas (Springer-Verlag, Berlin, 1996). [astro-ph/9312012]

\bibitem{Banff}
B. L. Hu, in {\it Proceedings of the Third International Workshop
on Thermal Fields and its Applications}, edited by R. Kobes and
G. Kunstatter (World Scientific, Singapore, 1994)[gr-qc/9403061]

\bibitem{CHM}
E. Calzetta, B.-L. Hu, F. D. Mazzitelli, Phys. Rep. {\bf 352},
459-520 (2001).

\bibitem{RH1}
    S. A. Ramsey and B. L. Hu,
          Phys. Rev. D {\bf 56}, 678 (1997).

\bibitem{RH2}
    S. A. Ramsey, B. L. Hu and A. M. Stylianopoulos,
     Phys. Rev. D {\bf 57}, 6003 (1998)

\bibitem{RHA}
A. Raval, B.L. Hu, and J. Anglin, Phys. Rev. D {\bf 53},
7003-7019 (1996).

\bibitem{RHK}
A. Raval, B. L. Hu and D. Koks, Phys. Rev. {\bf D 55}, 4795
(1997).

\bibitem{KHMR}
     D. Koks, B. L. Hu, A. Matacz, and A. Raval,
     Phys. Rev. D {\bf 56}, 4905-4915 (1997).

\bibitem{KMH}
     D. Koks, A. Matacz, and B. L. Hu,
Phys. Rev. D {\bf 55}, 5917-5935 (1997)

\bibitem{CH00}
E. Calzetta and B. L. Hu, Phys. Rev. D {\bf 61}, 025012 (2000).

\bibitem{CHcorent}
E. Calzetta and B. L. Hu,
in preparation.

\bibitem{SCHR}
     G. J. Stephens, E. A. Calzetta, B. L. Hu, and S. A. Ramsey,
     Phys. Rev. D {\bf 59}, 045009 (1999)

\bibitem{CRVopensys}
E. Calzetta, A. Roura and E. Verdaguer, Physica A {\bf 319}, 188
(2003).

\bibitem{CRVtunnel}
E. Calzetta, A. Roura and E. Verdaguer, Phys. Rev. D {\bf 64},
105008 (2001); Phys. Rev. Lett. {\bf 88}, 010403 (2002)

\bibitem{RouVer00}
A. Roura and E. Verdaguer, Int. J. Theor. Phys.  {\bf 39}, 1831
(2000).

\bibitem{RouVer03a}
A. Roura and E. Verdaguer, in preparation.

\bibitem{HuShi}     B. L. Hu and K. Shiokawa,
     Phys. Rev. D {\bf 57}, 3474-3483 (1998)

\bibitem{Tomeso} K. Shiokawa,
 Phys. Rev. D {\bf 62}, 024002 (2000).


\bibitem{HRS} B. L. Hu, A. Raval and S. Sinha,
``Notes on Black Hole Fluctuations  and Backreaction" in {\it
Black Holes, Gravitational Radiation and the Universe: Essays in
honor  of C. V. Vishveshwara} eds. B. Iyer and B. Bhawal (Kluwer
Academic Publishers, Dordrecht, 1998) [gr-qc/9901010].

\bibitem{SRH} S. Sinha, Alpan Raval and B. L. Hu,
``Black Hole Fluctuations and Backreaction in Stochastic
Gravity'',  Found. Phys. 33 (2003) 37-64 [gr-qc/0210013].

\bibitem{GregBH}
G. Stephens and B. L. Hu, ``Notes on Black Hole Phase
Transitions", Int. J. Theor. Phys. 40 (2001) 2183-2200
[gr-qc/0102052]

\bibitem{GPY} D. J. Gross, M. J. Perry and L. G. Yaffe,
           Phys. Rev. D {\bf 25}, 330 (1982).

\bibitem{WhiYor} B. F. Whiting and J. W. York, Jr.,
  Phys. Rev. Lett. 61, 1336 (1988)

\bibitem{stogra}
B. L. Hu, ``Stochastic Gravity", Int. J. Theor. Phys. {\bf 38},
2987 (1999) [gr-qc/9902064].

\bibitem{HVErice}
B. L. Hu, E. Verdaguer, ``Recent Advances in Stochastic Gravity:
Theory and Issues" Erice Lectures May 2001, in {\it Advances in
the Interplay between Quantum and Gravity Physics} edited by P.
Bergmann and V. De Sabbata, (Kluwer, Dortrecht, 2002)
[gr-qc/0110092].

\bibitem{CQGrev} B. L. Hu and E. Verdaguer, ``Stochastic Gravity: A
Primer with Applications", Class. Quant. Grav. 20 (2003) R1-R42

\bibitem{LivRev} B. L. Hu and E. Verdaguer, ``Stochastic Gravity:
Theory and Applications", {\it Living Reviews in Relativity}
(2003)

\bibitem{RouVer03b} A. Roura and E. Verdaguer, in preparation.

\bibitem{KinQG} B. L. Hu, ``A Kinetic Theory Approach to Quantum Gravity'',
Int. J. Theor. Phys. (2002) [gr-qc/0204069]

\bibitem{DeW75}
B. S. DeWitt, Phys. Rep. 19, 295(1975).

\bibitem{BirDav} N. D. Birrell and P. C. W. Davies, {\it Quantum Fields
in Curved Space} (Cambridge University Press, Cambridge, 1982).

\bibitem{scg}
 E. Calzetta and B. L. Hu,
          Phys. Rev. D {\bf 35}, 495 (1987);
 A. Campos and E. Verdaguer,
          Phys. Rev. {\bf D49}, 1861 (1994).

\bibitem{meso}
B. L. Hu, ``Semiclassical Gravity and Mesoscopic Physics" in {\it
Quantum Classical  Correspondence} eds. D. S. Feng and B. L. Hu
(International Press, Boston, 1997) [gr-qc/9511077].

\bibitem {GRhydro}
B. L. Hu, ``General Relativity as Geometro-Hydrodynamics"
Invited talk at the Second Sakharov Conference, Moscow, May, 1996
[ gr-qc/9607070].

\bibitem{if}
    R. Feynman and F. Vernon,
          Ann. Phys. (NY) {\bf 24}, 118 (1963);
    R. Feynman and A. Hibbs,
          {\it Quantum Mechanics and Path Integrals},
          (McGraw - Hill, New York, 1965).

\bibitem{qbm}
A. O. Caldeira and A. J. Leggett, Physica {\bf 121A}, 587 (1983);
Ann. Phys. (NY) {\bf 149}, 374 (1983). H. Grabert, P. Schramm and
G. L. Ingold, Phys. Rep. {\bf 168}, 115 (1988). B. L. Hu, J. P.
Paz and Y. Zhang, Phys. Rev. {\bf D45}, 2843 (1992); {\bf D47},
1576 (1993).

\bibitem{HM2}
B. L. Hu and A. Matacz, Phys. Rev. D49, 6612(1994)

\bibitem{ctp}
    J. Schwinger,          J. Math. Phys. {\bf 2} (1961) 407;
    L. V. Keldysh,          Zh. Eksp. Teor. Fiz. {\bf 47 }, 1515 (1964)
          [Engl. trans. Sov. Phys. JEPT {\bf 20}, 1018 (1965)]
    R. D. Jordan,          Phys. Rev. {\bf D33 }, 444 (1986);
    E. Calzetta and B. L. Hu,
        Phys. Rev. D {\bf 35}, 495 (1987);{\bf 37}, 2878 (1988); {\bf D40}, 656 (1989).


\bibitem{ELE}
E. Calzetta and B. L. Hu, Phys. Rev. D49, 6636(1994).  B. L. Hu
and A. Matacz, Phys. Rev. {\bf D51}, 1577 (1995). B. L. Hu and S.
Sinha,  Phys. Rev. {\bf D51}, 1587 (1995). A. Campos and E.
Verdaguer, Phys. Rev. D {\bf 53}, 1927 (1996).

\bibitem{Parker} L. Parker, Phys. Rev. {\bf 183}, 1057 (1969).
\bibitem{Hawking} S. W. Hawking, Comm. Math. Phys. {\bf 43}, 199 (1975).

\bibitem {infcos}
 A. H. Guth, Phys. Rev. D 23, 347 (1981).
 A. Albrecht and P. J. Steinhardt, Phys. Rev. Lett. 48, 1220 (1982).
 A. D. Linde, Phys. Lett. 114B, 431 (1982). Phys. Lett. 162B, 281 (1985)

\bibitem{KolTur90} E.W. Kolb and M. Turner, {\it The early Universe}
(Addison-Wesley, Reading, Massachusetts, 1990); T. Padmanabhan,
{\it Structure formation} (Cambridge University Press, Cambridge,
England, 1993).

\bibitem{Linde90} A. Linde, {\it Particle physics and inflationary
cosmology} (Harwood Academic Publishers, Switzerland, 1990).

\bibitem {HuPhysica}
B. L. Hu, Physica  A158, 399 (1979).

\bibitem{MV0} R. Martin and E. Verdaguer,
Phys. Lett. B {\bf 465}, 113 (1999).

\bibitem{MV1}
R. Martin and E. Verdaguer,  Phys. Rev. D {\bf 60}, 084008 (1999).

\bibitem{MV2}
R. Martin and E. Verdaguer,  Phys. Rev. D {\bf 61}, 124024 (2000).

\bibitem{PH97} N. G. Phillips and B. L. Hu, Phys. Rev. D {\bf 55}, 6132 (1997).

\bibitem{HP0}
B. L. Hu and   N. G. Phillips,  Int. J. Theor. Phys. {\bf 39},
1817 (2000) [gr-qc/0004006].

\bibitem{PH1}
N. G.  Phillips and  B. L. Hu, Phys. Rev. D {\bf 62}, 084017
(2000).

\bibitem{PH2}
N. G.  Phillips and  B. L. Hu, Phys. Rev. D {\bf 63},  104001
(2001).

\bibitem{MTW} C. W. Misner, K. S. Thorne and J. A. Wheeler,
{\it Gravitation} (Freeman, San Francisco, 1973).

\bibitem{bhbkr}
J. M. Bardeen, Phys. Rev. Lett. 46, 382 (1981); P. Hajicek and W.
Israel, Phys. Lett. 80A, 9 (1980).

\bibitem {York}
    J. W. York, Jr.,
          Phys. Rev. D {\bf 28}, 2929 (1983);
          D {\bf 31}, 775 (1985); D {\bf 33}, 2092 (1986).

\bibitem{CamHu}
A. Campos and B. L. Hu, Phys. Rev. D58, 125021(1998); Int. J.
Theor. Phys. 38, 1253(1999)

\bibitem{Page82}  D. N. Page, Phys. Rev. {\bf D25}, 1499 (1982).

\bibitem{Smoot92} G.F. Smoot {\it et al.}, Astrophys. J. Lett.
{\bf 396}, L1(1992).

\bibitem{Pad93}
T. Padmanabhan, {\it Structure Formation} (Cambridge University
Press, Cambridge, England, 1993).

\bibitem{MukFelBra92} V.F. Mukhanov, H.A. Feldman and R.H.
Brandenberger, Phys. Rep. {\bf 215}, 203 (1992).

\bibitem{Bar80} J. M. Bardeen, Phys. Rev. D {\bf 22}, 1882 (1980).

\bibitem{Roura99} A. Roura and E. Verdaguer, Int. J. Theor. Phys.
{\bf 38}, 3123 (1999).

\bibitem{strforRenegades}
E. Calzetta and B. L. Hu, Phys. Rev. D {\bf 52}, 6770 (1995); A.
Matacz, Phys. Rev. D {\bf 55}, {1860} (1997); {\bf 56}, 1836
(1997); E. Calzetta and S. Gonorazky,  Phys. Rev. D {\bf 55},
{1812} (1997)


\end{thebibliography}


\end{document}